# Recommended Actions for the American Astronomical Society: CSWA's Perspective on Steps for a more Inclusive Astronomy - I. Background and Methods

Rachel Wexler[1] Patricia Knezek[1] Gregory Rudnick[2] Nicolle Zellner[3] Kathleen Eckert[1] JoEllen McBride[4] Maria Patterson[1] Christina Richey[5] Committee on the Status of Women in Astronomy

[1]Committee on the Status of Women in Astronomy, [2]University of Kansas, [3]Albion College, [4]University of Pennsylvania, [5]Jet Propulsion Laboratory/California Institute of Technology

**ABSTRACT**

In a series of two papers, we provide a comprehensive agenda of actions the American Astronomical Society (AAS) can take to create a more diverse and inclusive professional system for astronomers, with a focus on women astronomers. This first paper of the series outlines the background and methods, while the recommendations are treated in the second companion paper (Paper II). We take the stance that since the 2020 Decadal Survey (Astro2020) was delivered in 2021, with its first-ever set of recommendations on the State of the Profession, now is the time for the AAS to take decisive action to transform astronomy into a diverse and inclusive profession. In the spring of 2019, the CSWA surveyed the astronomical community to assess the popularity and feasibility of actions that the AAS can take to reduce harassment and advance career development for women in astronomy. Here we present the quantitative results of that survey and a synopsis of the free response sections, which are publicly accessible. By combining the results of our survey, peer-reviewed academic literature, and findings from many of the white papers submitted to Astro2020, the CSWA has developed 26 specific actions that the AAS can take to help end harassment in astronomy, to advance career development for astronomers who are women and who are other members of historically marginalized groups, and intersections of these populations, and to improve the climate and culture of AAS and AAS-sponsored meetings. This paper presents the data we used to make these recommendations, and the recommendations themselves will be presented in Paper II.

# 1. Introduction

The American Astronomical Society's (AAS) Committee for the Status of Women in Astronomy (CSWA) was created in 1979 with a mandate to review the status of women in astronomy and make practical recommendations to the AAS Council (now the AAS Board of Trustees) on actions to advance the status of women in astronomy. In the following, the CSWA interprets "women" to mean people who identify as female, including trans women, genderqueer women, and non-binary people who are significantly female-identified (Knezek et al. 2020). As of today, the CSWA provides guidance to a broad range of stakeholders beyond the AAS, from individual astronomers to the federal government.

The CSWA has undertaken projects to facilitate wider communication and networks of support among women and minority astronomers and has been successful in creating awareness of the issues for women in astronomy. Through the Women in Astronomy meetings, regular sessions at AAS meetings, the *AASWomen* newsletter, and the *Women in Astronomy* blog, the CSWA has grown its ability to communicate internally and externally about the needs of women in the profession. The CSWA acts as a set of eyes and ears, and as a voice for women to help make positive changes in the field (Schmelz, 2011).

Most recently, the CSWA surveyed the astronomy community [3] to learn what specific areas and measures the astronomy community feels the committee should prioritize. These responses formed the foundation for the CSWA's Strategic Plan for the 2020s, which lays out the objectives that will guide the CSWA's activities for the next decade. This paper (hereafter Paper I) is the first in a two-paper series in which we present a series of recommendations to the AAS. Paper I presents the background and methods used to derive our recommendations. Paper II in the series presents those recommendations which align with the strategic plan and support the CSWA's progress towards its mission to build an inclusive and self-sustaining community that supports gender equity and the success of women in astronomy.

Social norms, the scientific profession, and astronomy have transformed in the 43 years since the CSWA was created. It is no longer enough to target only sexist discrimination in education and in the workplace. We choose to critically examine astronomy using an intersectional approach, meaning that we seek to address the needs of astronomers from all historically marginalized groups[1]. Our focus is on the experiences of women, and we seek to advocate for *all* women, including those who are LGBTQPAN, non-White, and/or disabled[2]. LGBTQPAN is an acronym referring to lesbian, gay, bisexual, transgender, queer/questioning, pansexual, asexual, and/or nonbinary (LGBTQPAN) people (Richey et al., 2020).

It is also no longer enough to simply bring more women and other members of historically marginalized groups into astronomy. While diversity is important, its full benefits will not be realized until all environments where astronomers work are free from discrimination and harassment and where institutions work proactively to make these environments inclusive. Women, and especially women of color in this field, report unacceptably high rates of harassment (Clancy, Lee, Rodgers, & Richey, 2017). As outlined in Paper II, central to our recommendations is a plan to strengthen the AAS's anti-harassment policies and procedures. As cohorts of graduating astronomers are increasingly diverse, the AAS must take action to support these scientists by reducing, and ultimately eliminating, harassment in all of its forms (National Science Foundation, 2018).

2022 was a key year for the astronomical community as we started to implement the recommendations of the Astro2020 Decadal Survey (National Academies of Sciences, Engineering, and Medicine, 2023). It is against this backdrop that we present this paper series. Our recommendations for the AAS, if enacted, will help create a strong foundation for positive change for the coming decade. Through a range of actions, some minor and some resource-intensive, the AAS can and should lead the effort to transform astronomy into a more diverse and inclusive profession in the coming decade and beyond.

This paper is organized as follows. In Section 2 we review the status of women in astronomy as a motivation for our continued efforts for the equitable, inclusive, and just treatment of astronomers across a range of different marginalized identities. In this section we consider the following topics: Participation (Section 2.1), Harassment and Bullying (Section 2.2), and The Impacts of Systemic Inequity and Discrimination (Section 2.3). In Section 3 we present the methods we used to gather our data. In Section 4 we present our synopses of the survey's free-response questions in four areas: Harassment & Bullying (Section 4.1), Creating Inclusive Environments (Section 4.2), Professional Development and Retention (Section 4.3), and Ethics (Section 4.4). We summarize our methods and background in Section 5. In the Appendix we present our quantitative data for the Likert-opinion scale questions. The free-response replies and a PDF of the original survey are provided in an [on-line repository(Zellner, Knezek, Wexler, & Committee on the Status of Women in Astronomy, 2022)](#).

## 2. The Status of Women in Astronomy

The number of women earning astronomy PhDs is slowly increasing. Women in astronomy, especially women of multiple minority identities, face challenges related to harassment and discrimination at unacceptably high rates. Women are still paid less than men and are disproportionately impacted by the two-body problem and caregiving responsibilities. In this section we outline the status of women in astronomy, focussing on three different areas: the [participation of women in astronomy](#)—including their share of degrees earned, the [effect of harassment and bullying on women in astronomy](#), and [the impacts of systemic inequity and discrimination.](#)

### 2.1 Participation

*The number of women earning astronomy PhDs each year is slowly increasing, along with the percentage of degrees earned by women compared to men. Women, especially women of color, are underrepresented in high-ranking positions compared to men.*

As reported by the National Science Foundation [(NCSES, 2022)](#), the number of women in astronomy has been continuously increasing over the past two decades ([Figure 1](#)). However, an examination of the trends in who earned degrees from 1998-2017 reveals that there has been only a modest increase in the proportion of women present in astronomy and astrophysics doctoral cohorts since the early 2000s ([Figure 2](#)).

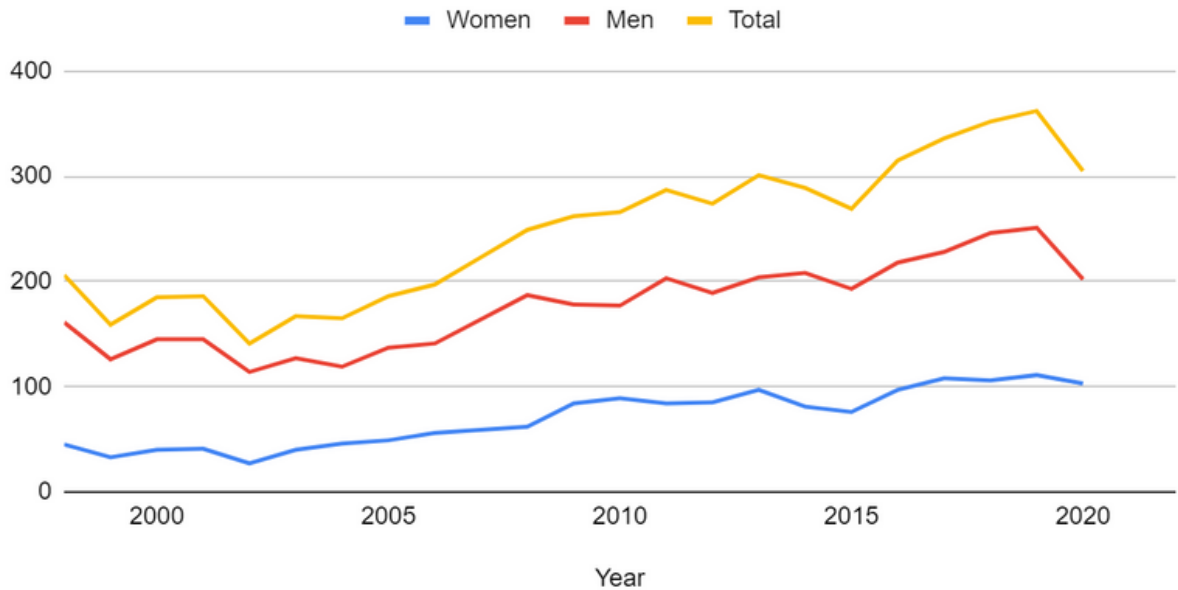

**Figure 1**
*Data from the NSF Survey of Earned Doctorates Data, showing the numbers of men, women, and total PhDs earned in Astronomy and Astrophysics between 1998 and 2020 (NCSES, 2022).*

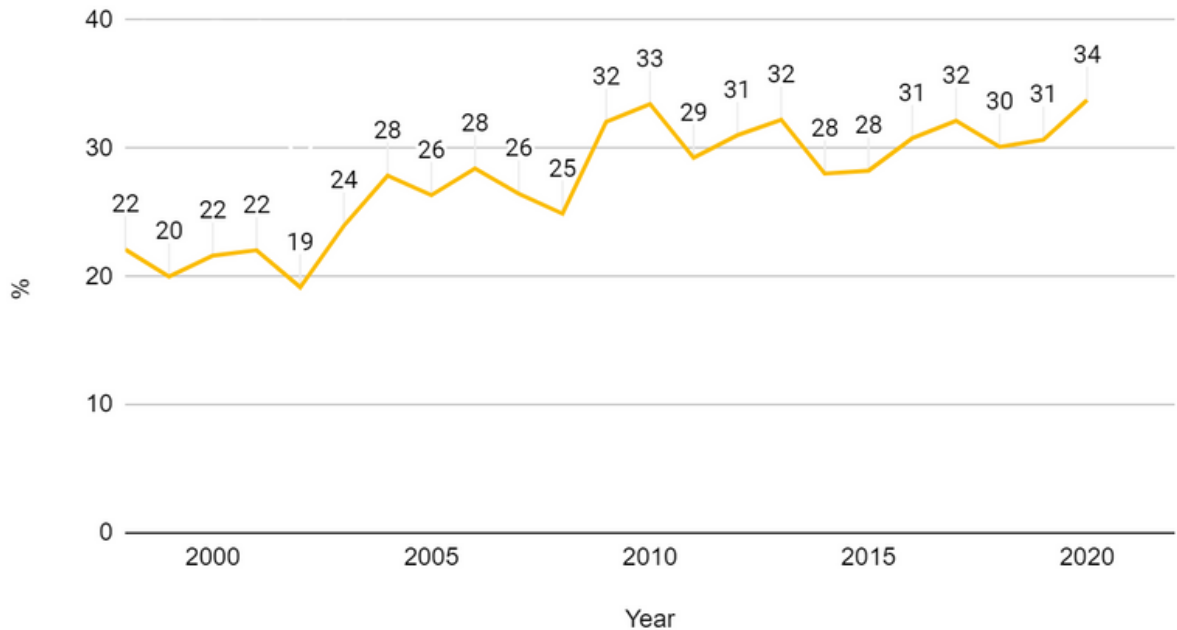

**Figure 2**
*Data from the NSF Survey of Earned Doctorates Data, showing the percentage of women who earned PhDs in Astronomy and Astrophysics between 1998 and 2020 (NCSES, 2022).*

In addition to gender imbalances, extreme disparities exist in doctoral degree earning for people of color compared to White women and men. In 2017, of the 336 earned astronomy doctorates (both genders), 9 were earned by Hispanic or Latinx scientists and 3 were earned by Black scientists. No doctoral degrees were awarded to American Indians or Alaska natives (National Science Foundation, 2018).

One Astro2020 white paper (Momcheva, 2019) illuminates the implications of the gender imbalance in astronomy. Momcheva (2019) measured women's participation in the field from 1970-2018 by counting the number of dissertations published in the Astrophysics Data System (ADS) each year and by tracking rates of publication over time by the dissertation authors. They found that overall, women represent about 19% of all PhD recipients over the last 48 years, starting at 5% in the 1970s and peaking at 30% in recent years. These data are consistent with the trends reported by the NSF.

Using the authorship trends data, Momcheva (2019) estimated that about 50% of the women she identified have dropped out of active research. These data are consistent with studies that suggest women drop out of STEM employment (including industry and academia) at high rates. For example, a recent study found that close to half of women in STEM who become mothers leave full-time STEM employment after having their first child and do not return (Cech & Blair-Loy, 2019). The historically low representation of women in past PhD-earning cohorts coupled with womens' high attrition rate compared to men is one of the reasons that the total number of senior women in astronomy is low today (Momcheva, 2019).

From their 2018 Academic Workforce Survey, the American Institute of Physics (AIP) determined that about 23% of astronomy faculty are women. More women are in lower ranks (assistant professor) than in higher ranks (full professor). Among recent astronomy PhD earners, women may be entering faculty positions at higher rates than their male counterparts, but are more likely to be offered research or adjunct faculty positions than tenure-track (assistant professor) positions. In 2016, women made up 32% of new non-tenure-track hires in astronomy, but only 23% of new tenure-track hires. This is consistent with research that has found that across STEM fields, women tend to be overrepresented as instructors and lecturers, and underrepresented in tenure-track positions (Fox et al., 2017; Porter & Ivie, 2019).

One study of PhD recipients from 2000-2012 reported that women are more likely to successfully transition from postdoctoral positions to junior faculty positions compared to men (Perley, 2019). However, we take the stance that hiring disparities between men and women are still present, as suggested by the AIP's findings and by cross-field studies that have found that women are less likely to obtain high-ranking positions, even when taking time since PhD into account (Fox et al., 2017). The long-term historical inequities that have advantaged White men, and disadvantaged all others, have long-reaching ramifications that leave their imprint on the representation of women in senior positions.

Further, the impacts of historical hiring discrimination against women continue to be felt. For example, The Maunakea Gender Equity and Diversity Committee published the results of a survey deployed to all Maunakea observatories in 2018; it found that 44% of women in STEM roles on Maunakea feel that their gender will make it hard for them to advance in the future (Dempsey, 2018).

It is also clear that greater disparities persist for women of color, especially Black women, compared to White women. White women, Hispanic women, and Asian women are entering faculty positions in increasing numbers, while Black women are not. From 2008-2016, the number of White women astronomy faculty increased from 465 to 661, while the number of Black women faculty remained the same, at 14 (Porter & Ivie, 2019). Over the same period, the number of Hispanic women faculty increased from 19 to 33, and the number of Asian women faculty increased from 106 to 185.

The low number of women, especially Black women, who are associate and full professors is a key concern. These senior women are sought after to mentor the growing numbers of upcoming women *and* are expected to participate in other forms of community service at their academic institutions, while still conducting research and being active members of their professional communities. This additional workload, often not shared by White, cisgender, male colleagues, contributes to systemic inequities in the workplace. While community service is found to be important to women and other members of historically marginalized groups, it is also important that *everyone* shares the work of supporting these scientists, especially because this kind of labor is typically not compensated (Ko, Kachchaf, Ong, & Hodari, 2013).

Women are still excluded from some key positions in astronomy, as demonstrated by a few case studies. A study conducted by Julie Rathbun showed that the percentage of women involved in planetary spacecraft science teams has remained flat, at 15.8%, from 2001-2016, even while the number of women in the field has continued to increase [18]. Rathbun (2016) stated:

> *While the percentage of women participating in astronomy and planetary science appears to be increasing, their representation on spacecraft science teams has not been commensurate. For many planetary scientists, being involved in a spacecraft mission is the highlight of a career.*

Another study found that of the 108 Astrophysics Explorer-Class mission proposals submitted to NASA from 2008-2016, only 5% of proposing Principal Investigators (PIs) were women. The median number of women scientists on these proposing teams was 3.5, and 18 teams had no women in science roles at all (Centrella et al., 2019).

We seek to support the attraction, retention, and promotion of all women, including gender non-conforming scientists. Unfortunately, little data exist on LGBTQPAN participation in physics and astronomy. Even so, our survey respondents expressed a desire (see Section 4) to see more support for LGBTQPAN scientists, in addition to women scientists and scientists of color. The NSF plans to pilot questions regarding gender identity and sexual orientation on its workforce surveys, so in the future, these data will likely exist [21].

Astronomy has reached a critical point for women. The rate at which women enter the field is slowly increasing (Figure 1, Figure 2), but hiring and retention has not reached gender equality (Kewley, 2021; Feder, 2022). It is also clear that astronomy is not attracting, retaining, and promoting women of color at nearly the same rates as White men and women (Porter & Ivie, 2019).

Even so, as the field grows in overall population, more women than ever are earning PhDs in astronomy and astrophysics (Figure 1), although the percentage of PhD recipients who are women has stagnated over the last 10 years (Figure 2). Nonetheless, the AAS has more women to serve and must take action in the near future to shape women's experiences for the better, so that they can achieve their full scientific potential. Multiple studies have shown that diverse teams are more creative, and are better at solving problems (Hong & Page, 2004; Phillips, 2022). The CSWA website features a full set of resources that detail the benefits of diversity (CSWA, 2019). To give one example, when interviewed, one Black woman graduate student in physics explained how her intersectional identities make her a better scientist (Chambers, 2019):

> *A lot of the training that you get in physics is reductionism. You take a complex system and you make it very simple...Whereas when you're African American and/or a woman, when you're intersectional, you have to deal with complexity all the time... And so that means that the way that I dealt with my*

> *dissertation, I looked at something that was complex and I tried not to reduce the complexity; I tried to weave the complexity together.*

Diversity is important to science because diversity *fundamentally changes the questions that are asked in science*. When diversity is absent, ideas are left unexplored. By 2045, the United States population will be over 50% non-White [(Frey, 2018)](). We need to address barriers to success for all scientists, especially women and other members of historically marginalized groups, because it is the moral way forward, because better institutional climates and more cogent teams produce better results, and because diversity leads to more creative research.

## 2.2 Harassment and Bullying

*Harassment and bullying are prevalent for women in the sciences, including women in astronomy. Women of color and LGBTQPAN women are even more likely to be targets of harassment compared to White, cisgender women.*

In 2018, the National Academies of Science, Engineering, and Medicine published a report titled *Sexual Harassment of Women: Climate, Culture, and Consequences in Academic Science, Engineering, and Medicine* (referred to hereafter as "the NASEM report") When. The NASEM report reviews the best available research on sexual harassment in the sciences, and includes findings from an original cross-field study of the experiences of 40 women scientists. NASEM's cross-field interview study found that harassing behavior is learned intergenerationally. When younger scientists observe older scientists engaging in harassing behavior, they become more likely to harass. Underrepresentation of women and minorities correlates with increased harassment; and organizations led by diverse teams see lower rates of harassment. Therefore, increasing the number of women and minorities in the field, and in leadership roles, is an important component of a holistic plan to address harassment.

Women often feel the *intellectual impacts of harassment*. After experiencing harassment and bullying, women interviewed as part of the NASEM study reported that they felt less inclined to share their ideas and input. It is important to note that environments where harassment exists are bad for everyone, regardless of gender or racial and sexual identity. Harassment and bullying that occur within a work environment, but do not directly impact every individual (ambient harassment), contributes to the collective stress of teams and decreases creativity and knowledge flow [(NASEM, 2018)]().

The patterns of harassment in academia and science are consistent with the few studies that are specific to astronomy. Dr. Kathryn Clancy's research team conducted a survey of astronomers and planetary scientists to assess workplace conditions for women in the space sciences, with a focus on women of color. 30% of female respondents reported feeling unsafe in the workplace because of their gender, and 40% of women of color reported feeling unsafe in the workplace because of their race or gender. Most notably, 18% of women of color and 12% of White women reported having skipped professional events in the past because they felt unsafe attending them (Clancy, Lee, Rodgers, & Richey, 2017).

An inhospitable climate need not be the result of severe instances of harassment or bullying. Microaggressions can also play a significant role in negatively impacting the climate and the degree to which women and other members of historically marginalized groups feel welcome and are included. The impacts of inhospitable climates of meetings and conferences on women are reflected by two surveys of conference attendees, one finding that women astronomers and planetary scientists were less likely to ask questions compared to men at the 223rd AAS meeting [31], and another finding the same at the Survey of Lyot conference in 2019 (Lebelloux et al., 2020). Notably, the study of attendees of the Survey of Lyot conference found that women were more than twice as likely to report having been interrupted while talking at a professional event compared to men, and far more likely to feel that they had been treated differently at a professional event because of their gender (Lebelloux et al., 2020). These studies point to a loss of career opportunities and suppressed professional participation of women because of both acute and ambient harassment.

In 2016, the American Physical Society published the LGBT Climate in Physics Report. Their report included the results of a survey to which 15% of LGBT male scientists, 25% of LGBT female scientists, and 30% of gender-nonconforming scientists responded that the overall climate of their department or division is "uncomfortable" or "very uncomfortable." In-depth interviews with LGBTQPAN physicists found that LGBTQPAN physicists of color face more harassing comments than their White counterparts. LGBTQPAN physicists also struggle to identify allies in their workplaces who are willing to support them (Atherton et al. 2016). An analysis of the Clancy team's survey of planetary scientists and astronomers found that LGBTQPAN respondents were more likely to respond that they feel unsafe at work, and LGBTQPAN women were more likely to report being verbally and physically harassed compared to cisgender, straight women [4].

These results reflect a phenomenon we want to emphasize: *Women and gender non-comforming scientists with multiple minority identities experience more discrimination, harassment and bullying than do men and white women.*

A different survey conducted by Clancy's team collected responses from 666 scientists who work at field sites, mostly in the biological sciences. 64% of scientists of all genders working at field sites reported having been harassed, and *20% reported having been assaulted*. Notably, most perpetrators had a superior rank to their targets (Clancy, Nelson, Rutherford, & Hinde, 2014). Off-campus work is important to consider in astronomy as well, as astronomers often work at off-campus telescope sites.

Across professional organizations, federal funding agencies, universities, and observatories, harassment reporting policies and procedures vary. It is difficult to know the rate at which harassment goes unreported; however, it is expected to be high. For example, 66% of female respondents to the Maunakea survey said they would be uncomfortable reporting harassment (Dempsey, 2018).

The NASEM report hypothesized that women and LGBTQPAN scientists are more likely to be harassed because they don't conform to typical "scientific" expectations of appearance and behavior (NASEM, 2018). For example, in a sociocultural study examining the experiences of two astrophysics doctoral students, a student reported being purposefully ignored by older male colleagues after she became pregnant, because, in her own words, pregnancy made her appear more "visibly female" (Gonsalves, 2018).

Federal funding agencies and other professional societies have taken action to improve their ability to sanction those who are found to be perpetrators of harassment. The NSF has implemented a new term and condition that mandates NSF-funded institutions report to NSF when a funded Principal Investigator (PI) or co-PI is found guilty of harassment or assault, or when action is taken against them pending an investigation of accusations thereof. The NSF has stated that upon receiving such information, its "actions may include substituting or removing principal investigators or co-principal investigators, reducing award funding, and -- where neither of those options is available or adequate -- suspending or terminating awards." (NSF, 2018). NASA implemented a similar term and condition for its grants in March, 2020 (NASA, 2020). The American Geophysical Union (AGU) has taken a strong stance against harassment and bullying by including harassment and bullying in its definition of research misconduct, making it so that harassment and bullying are considered to be as serious as offenses such as falsifying research results (American Geophysical Union. n.d.). While the AAS has not done this, it has elevated Harassment and Bullying to the same level as Research Misconduct in its Code of Ethics.

## 2.3 Impacts of Systemic Inequity and Discrimination

*Systemic inequity, discrimination, and stereotypes about women and science create negative outcomes for women, including less pay, less recognition, and other barriers to career success.*

In addition to being more likely to face harassment, women are more likely to be underpaid and face barriers to career advancement. For example, women are less likely to receive telescope time compared to their male colleagues (e.g., Kuo, 2016; Moskowitz, 2014). Women are also disproportionately impacted by caregiving responsibilities and the two-body problem (e.g., Finkel & Harvey-Smith (2020) and references therein), and most recently, the COVID worldwide pandemic (e.g., Myers et al., 2020; Squazzoni et al., 2021).

### 2.3.1 Compensation

Across scientific fields, women's salaries are 11% lower than men's salaries. According to the NSF's 2021 Survey of Doctorate Recipients (NCSES, 2023; their Table 59), women in astronomy and astrophysics earn a median of $83,000 per year, while men earn a median of $90,000 per year. The same survey found that women full professors in astronomy and astrophysics earn a median of $113,000 per year, while men full professors earn a median of $128,000 per year. However, women associate professors in astronomy and astrophysics earn a median of $103,000 per year, while men associate professors earn a median of $94,000 per year and women assistant professors were found to earn a median of $87,000 per year, while men assistant professors were found to earn a median of $77,000 per year.

The sources of these disparities are unclear. Men may be paid more overall and at the full professor stage because they make up the majority of more senior cohorts, are more likely to hold high-ranking positions, and are more likely to negotiate for higher salaries (Leibbrandt & List, 2015). Women full professors may earn less due to the career pauses necessitated by caregiving responsibilities, which will be further discussed in section 2.3.4. The larger salary earned by women at the associate and assistant professor level is interesting, although we do not have any conclusive explanation at the present time. These data suggest that institutions employing astronomers are beginning to critically examine pay structures but also indicate that there are still improvements to be made, especially at the highest levels of rank (Fox et al., 2017; NCSES, 2023). Our recommendations to improve compensation for women and other members of historically marginalized groups in astronomy aim to increase transparency about pay structures, because transparency about compensation leads to increases in pay equity and job satisfaction (Hartmann & Slapničar, 2012; Ramachandran, 2011).

Graduate students and recent PhD recipients in science are frequently underpaid compared to those who choose to work in industry after earning undergraduate degrees. Little recent data are available on compensation for graduate students in astronomy. However, in 2010, the AIP reported that first-year physics graduate students were typically paid between $10,000 and $25,000 per academic year (AIP, 2010). In the same year, the median salary for physics and astronomy bachelor's earners was $75,000. According to the NSF's 2017 Survey of Earned Doctorates, the average postdoctoral researcher (postdoc) in the physical sciences was paid $46,000 per year in 2017 (NCSES, 2018), but salary is inconsistent across subdisciplines and genders. A recent cross-field survey of postdoctoral researchers shows they may be paid anywhere from $23,660 (the U.S. minimum wage) to over $100,000 per year, and that male researchers are paid approximately $1,700 more on average than female researchers (Athanasiadou, Bankston, Carlisle, Niziolek, & McDowell, 2018).

As indicated by the above data, astronomy may fail to attract and retain women, especially women of color, because of the financial difficulties scientists experience during graduate school and postdoctoral work. A mixed-methods study (Ko, Kachchaf, Ong, & Hodari, 2013) of women of color in physics and astronomy (both in academia and in industry) found that women of color are more likely to come from low-income families and may need to provide for parents and siblings, possibly in addition to their own spouses and children. These women often choose industry over graduate school or postdoctoral work because industry physicists are paid better in the years immediately following bachelor's degree completion (Ko, Kachchaf, Ong, & Hodari, 2013). A study using findings from the Longitudinal Survey of Astronomy Graduate Students (LSAGS) found that astronomers who are members of historically marginalized groups were more likely than astronomers who aren't to seek work outside astronomy because of inadequate compensation (Ivie et al., 2017). Improving pay and transparency around pay are key to increasing the number of women and women of color in astronomy.

Still, significant gaps exist in our understanding of how the lack of adequate compensation for graduate students and postdocs impacts the career decisions of women and other members of minority groups. Payment schemes (including healthcare benefits) for graduate school vary greatly across institutions and regions, and standards for paying graduate students do not exist. Poor compensation disproportionately impacts students with health issues, those from low-income backgrounds, and anyone with family caregiving responsibilities.

### 2.3.2 Allocation of Telescope Time

Three studies examining telescope time reveal that women astronomers continue to be impacted by implicit gender biases. Reid (2014) found that male Principal Investigators (PIs) had a higher success rate when requesting Hubble time for cycles 11-21. He reported that this pattern suggests a systemic problem. Lonsdale, Schwab, & Hunt, (2016) found that there is a gender bias reflected by proposal selection patterns for the four NRAO facilities, and for ALMA, and Patat (2016) and Kuo (2016) both found a gender bias in time allocation for the ESO telescope. Taken together, these studies reveal that astronomical selection processes may benefit from further measures to mitigate the impacts of implicit bias, such as those taken by the Space Telescope Science Institute (e.g., dual-anonymous peer review), which have improved the equitable allocation of Hubble telescope time (Aloisi & Reid, 2021; Johnson & Kirk, 2020).

### 2.3.3 The Two-Body Problem

The two-body problem refers to the issues that arise when a couple struggles to find jobs within commuting distance of each other. It is common in academia. In astronomy, the two-body problem disproportionately impacts women because women in astronomy are more likely to have a partner who is also an astronomer. Specifically, women in physics and astronomy are 204% more likely than men to relocate for a spouse, and 346% more likely than men to turn down a job for a spouse (Porter & Ivie, 2019). The two-body problem is especially salient for early career academic couples, who often need to relocate to pursue graduate school, postdoctoral fellowships, and professorships.

One study used data from the Astronomy Rumor Mill to examine astronomers' career progressions, and found that women may be likely to leave astronomy after completing postdoctoral work due to pressures associated with the two-body problem (Flaherty, 2018). The two-body problem is prevalent, and attention and resources are needed to address it.

### 2.3.4 Caregiving

Women are more likely than men to take on caregiving responsibilities, including both childcare and eldercare. A study using data from the NSF's Scientists and Engineers Statistical Data System shows that 43% of women, but only 23% of men, in STEM leave full-time employment after having their first child (Cech & Blair-Loy, 2019). There are no data on how eldercare impacts the careers of scientists; however, the Family Caregiver Alliance reported that overall, 65% of family caregivers for older adults are women (Family Caregiver Alliance, 2019). Policy changes and resources (e.g., grants, flexible leave) to support caregivers will benefit caregivers of all genders, especially women.

The impacts of the two-body problem and caregiver issues are exacerbated by destructive social norms about the nature of astronomical work. Sociological studies of astronomy reveal that within universities, female-identifying astronomers are often subject to settings where their peers and superiors, often male, either do not seem concerned with work-life balance, or can easily achieve work-life balance because of the support of a stay-at-home partner (Barthelemy et al. 2015). These norms may make academic science unattractive to many who may have been excellent scientists and could be harmful to the mental health of women and other members of historically marginalized groups as they persist throughout their careers.

## 3. Methods

Data on participation in astronomy come from reports released yearly by the National Science Foundation (NSF), including the Survey of Earned Doctorates and the Survey of Doctorate Recipients. Our recommendations to decrease harassment in astronomy are informed by the National Academies of Science, Engineering, and Medicine's report on the Sexual Harassment of Women, published in 2018. We also benefit from continuous long-term discussions with the AAS sister inclusion committees: the Committee on the Status of Minorities in Astronomy (CSMA), the Committee for Sexual and Gender Minorities in Astronomy (SGMA), and the Working Group on Accessibility and Disability (WGAD).

Data collected by the CSWA inform this paper. In 2019, the CSWA conducted a survey (Zellner, Knezek, Wexler, & Committee on the Status of Women in Astronomy, 2024) to better understand the concerns and preferences of the astronomical community. To inform the content of the survey, four total sessions were held, two at the winter 2018 AAS meeting and two at the summer 2018 AAS meeting, to discuss the main issues of importance to astronomers who identify as women. The survey was designed in late 2018 and opened in early 2019. It was advertised through a variety of means including: a Women in Astronomy Blog post, the AAS News Digest, and on social media. Our survey assessed astronomers' perspectives on policies in four areas of concern: 1) harassment and bullying, 2) creating inclusive environments, 3) professional development, and 4) ethics. The survey included 53 Likert-scale questions (i.e., questions with a multi-point scale that lets the individual express how much they agree or disagree with a particular statement) that allowed astronomers to rate the effectiveness of policy actions that could be undertaken by AAS, and 17 free-response opportunities to explain answers and introduce new ideas.

We received over 340 responses to the survey. No personally identifiable information was collected and, as a result, we cannot report a breakdown of responses by gender, race, sexual identity, gender identity, career level, institution, or other characteristics. We acknowledge the limitations of being unable to examine correlations between respondents' identities, their professional ranks, and their preferences. Although such data would have been informative, we believe the anonymous nature of the survey increased astronomers' willingness to take the survey and write candid free responses. Note that there were multiple comments throughout the survey that criticized the shortcomings of the survey and highlighted the need for outside experts to aid the CSWA and AAS in identifying and tackling these issues. The authors of this survey completely agree and call upon the AAS to provide resources to bring in outside experts. The CSWA has additionally emphasized the use of outside experts in their Strategic Plan for the 2020s, and certain efforts related to this plan are currently being executed in cooperation with social scientists.

The call for community input to help guide the 2020 Decadal Survey (Astro2020) included a call for white papers on "Activities, Projects, or State of the Profession Considerations." A total of 312 white papers were submitted, and close to 30 of these white papers focused on or had strong connections to issues of harassment, diversity, and inclusion. These papers (Zellner et al. 2019a, 2019b) also inform our recommendations.

## 4. Data

As described above, the questions in the survey were composed of Likert-scale questions as well a space for free-response comments. These free response comments are provided as an online dataset at this link (Zellner, Knezek, Wexler, & Committee on the Status of Women in Astronomy, 2022). In the appendix we provide the full breakdown of the answers to the Likert-scale questions. We took all of the free response questions and, within each focus area of the survey (Harassment & Bullying, Creating Inclusive Environments, Professional development, hiring, and retention, and Ethics), we group comments thematically based on the content of the response. Below, we summarize the themes from each focus area and give example responses. To best understand the context of the comments below, it is useful to look at the questions asked in the relevant Likert-scale sections. Since each subsubsection below refers to a specific group of questions, the title of that subsubsection is referring to feedback in addition to the Likert-scale section.

We have included representative quotes verbatim from the survey. All quotes are given in *italics*. In each subsection, we present the subsubsections in the order that they appeared in the survey, which is the same order as the Likert-scale questions in the appendix. For example, subsubsection titles such as 4.1.1. "Other steps to prevent harassment" are referring to a block of Likert-scale questions that provide different possible steps. Therefore, the "other steps" in this subsubsection was intended to give the reader an avenue to provide input not encapsulated in the Likert-scale questions.

### 4.1 Harassment & Bullying

#### 4.1.1 Other steps that the AAS could pursue in order to prevent harassment.

1. There were worries about legal overreach by the AAS and potential legal problems should an escrow policy be implemented
2. There were requests that institutions with a history of incidents be identified.
3. There was interest in releasing some kind of information about cases of harassment, but there was no consensus on whether it should be anonymized nor consensus on the degree to which information should be released and people held accountable. There was general support for protecting victims by keeping them anonymous.

> *I favor 'accountability' more than anonymized information, partly because anonymizing is difficult, and in some cases in conflict with accountability. Some perpetrators should be both held accountable AND given an opportunity to learn better/different behavior. In fact, there may be many of us who contribute to a harassing culture (by tolerating it, consciously or unconsciously) who need to be held to account, and who need to learn new behaviors, but who do not deserve to be punished.*

4. There were multiple comments in support of expanding harassment beyond gender harassment.

    > *Please include racial harassment in addition to sexual harassment training*

    > *upgrade mental disability to be equal to other disabilities.*

    > *Broaden considerations of harassment: lately it emphasizes harassment of women but (as exemplified by the examples in your definition above) it de-emphasizes age-discrimination, which is extremely prevalent in planetary science and astronomy.*

5. There were multiple comments in favor of banning harassers from meetings and removing their AAS membership

6. There were very mixed reviews on training. For example:

    > *Research overwhelmingly shows that bystander training is the only training that really works, yet it is rarely available. Would it be possible to hold such training during big conferences like the AAS and AGU? This not only gives an opportunity to people that do not have that option at their home institution but also sets an example that may be hard to ignore.*

    > *Studies show that training doesn't work, and probably makes harassment worse.*

7. There were many comments that directed action that needs to be taken by Universities, not the AAS.

    > *We need to enforce cross-institutional, cross country harassment policies. A european postdoc here, who is being harassed since years by her former phd supervisors. Despite my reports to my current institution of several current incidents, my employer says they cannot do anything because my harassers are not employees of my institution and reside in another european country. Moreover, my university says that I am not protected by them while i am on work trips abroad, or even outside the campus, because then different laws apply. They say they can only do something if I am being harassed on campus, or if outside then only when the harasser is an employee of the same university. I find it unacceptable, that when we are on work trips, we have zero rights.*

### 4.1.2 Other steps the AAS could undertake to support those who may be or have been harmed by harassment and bullying?

1. Comments mentioned that the AAS can offer counseling or legal advice at low or no cost, though there was concern as to the circumstances in which counseling would be covered and the degree to which the AAS could navigate specific institutional processes. It was brought up that the AGU provides free legal advice. Counseling and legal advice would need to be provided by qualified professionals. However, there were multiple respondents who worried that offering such services would be outside the scope of the AAS.

> *PLEASE consult experts in these areas. You wouldn't ask a non-expert to weigh in on the best way to measure stellar multiplicity. Don't ask astronomers the best way to mitigate harassment and bullying; seek out and pay experts for this work.*

2. There was a recognition of the limited power of the AAS to take action at the institutional level but also some desire for the AAS to put some pressure on institutions if harassment there is not addressed properly.
3. Comments expressed a desire to prevent retaliation, but there was no clear idea of how this should be done.
4. There was an emphasis on the role of people in power in the reactions to harassment. This can work in a negative way, in that people in power work to maintain the status quo and may not adequately address harassment issues. However, it can also be seen as a call to action for people in power to be more proactive in taking concrete steps to address systemic and individual issues of harassment.

> *Forcing behavior - even telling people to speak up for themselves, can often make the bullying and harassment worse. We have got to stop making the victims do double duty (receiving the aggression and fighting the aggression) and make leaders in these systems say "stop it" and hold their colleagues accountable. The leaders are the people in the power positions - they have the power to intervene. What I see is that they rarely ever do. The victims are powerless - they do not hold the power to stop the behavior or punish the aggressors.*

5. Comments reflected that it is important to not make the victim be the sole or main person to have to fight the harasser. This will require others, and those in power, to act in the victims' best interest.
6. There were multiple comments which indicated that harassers should be barred from meetings or have their AAS membership revoked. However, it was also recognized that the AAS has to clearly define in which circumstances it has authority to act. For example, the AAS would need to decide if they can bar someone from a meeting based on harassment that happened at another institution. This is similar to comments mentioned in 4.1.1.e, but we leave it here for completeness.

### 4.1.3 How should the CSWA and the AAS better coordinate with the AAS's other equity and inclusion committees (the Committee on the Status of Minorities in Astronomy [CSMA], the Committee for Sexual-Orientation & Gender Minorities in Astronomy [SGMA], and the Working Group on Accessibility and Disability [WGAD]) to address intersectional harassment?

1. There was a significant amount of support for joint meetings, perhaps held in dedicated sessions at the AAS.
2. Multiple respondents liked the idea of cross-committee memberships, though the authors of this report note that cross-committee memberships have the potential to overburden people from historically marginalized groups.

3. Multiple replies indicated that policies and training should be jointly addressed by all the diversity committees, with the goal of properly addressing intersectional issues.

   > *Any new training procedures should receive input from all the other equity and inclusion committees. Anti-harassment documents, emails, and events should have and show the support of all of these equity and inclusion committees. I think a united voice on harassment and bullying would be effective.*

4. There were strong statements about past CSWA actions that did not properly treat intersectionality and which were not LGBTQ friendly. These were often accompanied with warnings that the CSWA should embrace an intersectional approach in its future actions.

   > *For years the CSWA was solely focused on the topic of gender equality - but in doing so it forgot that people of color faced the same and even more severe conditions. In years past CSWA also focused primarily on issues affecting White and privileged women - again forgetting people of color. Similarly I recall there being some issues where CSWA did not act as a proper ally to the LGBTQ community's efforts at organizing within the AAS. I only talk about this history to warn the CSWA from becoming too insular, too driven by one or two strong personalities, too overly secretive and non-transparent and thus non-communicative with others joined in the same struggles.*

### 4.1.4 Should the AAS encourage or implement training at colleges, universities and/or other locations? Why or why not? Do you have any suggestions about how the AAS should do this or what the content of the training should be?

1. There was some support for the AAS to provide trainings. These comments pointed out that local institutions are more concerned with liability protection than actual prevention of harassment. However, at the same time there was also pushback to the AAS adding another layer of training on top of existing institutional training, and indeed sentiments were expressed that trainings in general were not useful.

   > *I think the AAS should develop effective harassment and bystander programs that can be sent to various institutions; these programs would need to reach beyond the "don't be illegal/minimizing liability" trainings that most departments currently offer. These trainings should include results and incorporate suggestions from the 2018 NASEM report on Sexual Harassment.*

2. Despite the mixed reviews of trainings overall, there was quite a bit of feedback highlighting the benefits of bystander training.

3. There were a large number of responses indicating that the AAS should encourage trainings at local institutions or work with those institutions to make effective and discipline-specific training.

   > *I think the best approach would be to develop training materials. My institution uses an online training now, but it's not very appropriate to academic astronomy departments. I think AAS would leave the legalistic part to the institution, but could provide the ethical part -- e.g. scenarios with gray zones that prompt discussion and contemplation.*

4. There were many comments suggesting that the AAS should maintain a list of institutions with mandatory training, perhaps with some kind of indicator as to whether these trainings were good or effective. This list would then be used by people when searching out institutions to which they would like to apply.
5. It was stated that trainings should be developed by qualified individuals.

   *I think AAS should definitely encourage this kind of training, but with the exception of a few people, we do not have the expertise within the society to do this well. I think AAS would be better off looking for experts in these areas, and sponsor them, rather than doing something poorly.*

6. There was overall skepticism at the utility of online training and more focus on the ability of AAS to help offer effective in person training.
7. There was interest for some training specific to astronomy/physics/planetary science situations.

   *It may be considered stepping beyond the boundary of AAS to impose something on all astronomy departments, but having a common, vetted training available to all universities with an astronomy department could be very useful. Consider making something available that 1) defines various terminology, 2) provides references, 3) identifies types of subtle micro-aggressions in addition to flagrant harassment, 4) informs about reporting, punishment, and policies at least within AAS if not with room to add university specific policies, and 5) gives scenarios applicable for common astronomy situations, like conferences, general meeting/team travel, observing runs in remote/isolated places with very few people, review panels, faculty promotion meetings, etc. AAS should lead by example. Departments could follow their lead.*

8. There was an idea of making mentoring workshops based on the idea of "ALMA days" where a trained ambassador hosts a workshop on mentoring. There was discussion of what kind of incentives AAS could build to have institutions participate in these workshops.

   *Absolutely yes. I think the key is to make these about antiharassment, but also about better mentoring and managing people. Professors are never trained in this, and that is partially a source of many biases and unwanted behaviors.*

   *One way to implement this could be similar to ""ALMA Days"", where a local ALMA ambassador runs a day-long workshop from materials prepared by ALMA. A huge incentive for institutions to have these happen could be to have institutions which do not participate be ineligible for all the AAS awards and grants.*

### 4.1.5 If you have experience with anti-harassment training, what method(s) have you found to be effective and not effective?

1. There were multiple comments highlighting the effectiveness of in-person training when compared to online training.

> *Both my home university and my current employer require anti-harrassment training. Between the two, I've found that the best training comes from people who's sole job it is to offer training coming from an agency with a lot of experience in it, and that training offered by a person or people is much better than training via videos or web-text content. An experienced training person will get everyone actively thinking. My current employer tried to offer training via a selection of videos, and some of those videos were objectively bad or had glaring problems or omissions, leaving several of us feeling uncomfortable or upset....*

2. There was widespread support for bystander intervention training. This is a repeat of a comment in 4.1.4b but we leave it here for completeness.
3. There were multiple comments highlighting that it is necessary to inform people what constitutes harassment, as there may be members of the community who are not aware that their actions are unacceptable.

> *As a former dean, I found that a very effective approach was to use peer pressure, i.e. the boys they go to lunch with. Often the offender didn't know they were doing anything wrong, or at least claimed they didn't*

### 4.1.6 Other comments on harassment and bullying

1. Respondents felt that bullying needed to be addressed, even if it did not rise to the legal definition of harassment. It was noted that bullying can happen between peers and between individuals who experience a power differential.
2. There were recurrent themes that called for a variety of actions against perpetrators of harassment and bullying. One of these comments was:

> *My impression is that harassment and bullying is much more prevalent in the sciences than one would think. A lot of the time it is a matter of abuse of power and a lot of the time the perpetrator is protected by their cohort of powerful peers. I have heard many times "oh they did not mean it this way" and "we have to be careful not to destroy this person's career". I rarely, even now, hear a genuine concern for what the victim is going through. This is a cycle that must be broken and it has to start with strong messages that this behaviour is absolutely wrong. For these messages to be strong they cannot be words alone but have to be followed with actions.*

3. It was noted in multiple places that bullying and harrassing did not always have a male as the perpetrator.
4. There were repeated comments indicating that enforcement is important, but that due process needed to be provided to the accused.

## 4.2 Creating Inclusive Environments

## 4.2.1 Comments on fostering equity and inclusion across the board, especially for community members with intersectional marginalized identities

1. It is important to publicize events and send out targeted invitations when appropriate.
2. The AAS should practice inclusion at its meetings (e.g., gender neutral bathrooms and childcare) and provide guidance on such practices to employment institutions.

   > *Along with making sure AAS meetings are accessible (at family-friendly times, flexible scheduling, video conferencing; in venues that are universally accessible), recommend all universities to do the same.*

   > *AAS should provide recommendations in support of reasonable benefits, especially for early-career researchers (grads and postdocs), including paid parental leave (or carer's leave for other family members) and medical leave, access to health insurance for families, access to proper mental health care coverage, etc.*

   > *The AAS needs to make sure that the salaries and grants received by under-represented groups are equal to those of white men in comparable positions. If women and others in under-represented groups have the same salaries and the same resources as those of men of similar stature, then they will command more respect and be less capable of being marginalized.*

3. Comments pointed out the need to minimize the burden on women and minorities through mentoring, awards, and encouraging people who have not belonged to historically marginalized groups to carry a share of the burden.

   > *While increasing equitable access to policy making and leadership roles is admirable, acknowledgement of the increased burden on women and minorities at higher levels of management must be acknowledged, and tokenism avoided. I think it is just as important to have a well oiled, diverse pipeline to the top as it to providing opportunities to those in minority positions into leadership roles at the moment. Our field is not diverse, but there is more than one way to improve things. Leadership and role models are important, they place significant burden on minority participants, who in turn end up falling behind. How can you avoid this? Perhaps a mentoring program could be effective? A near-peer ""buddy"" system? Increase awareness in how one rises through the ranks beyond getting tenure?*

4. There were multiple comments taking issue with the survey's inclusion of a question as to whether preventing reverse discrimination should be considered in the quest for inclusivity (Section A.2). The purpose of this question was not to endorse the idea of such a thing as "reverse discrimination", which is rightly considered by most people involved in DE&I efforts (and indeed the CSWA) to be an invalid descriptor of behavior. The goal of this question was instead to gauge the community's attitudes about the concept of "reverse discrimination", so as to understand what fraction of our community did believe that it is a valid issue. In that sense, answers mistakenly supporting the reality of "reverse discrimination" are a demonstration of the work that our community needs to do. While the quantitative results were evenly split across all categories, the free response answers all objected to the term even being used, as it implies "reverse discrimination" is a real phenomenon.

### 4.2.2 Other comments on promoting equity and inclusion

1. It was pointed out that there should be properly implemented gender neutral bathrooms at AAS meetings. It is especially important that women's bathrooms don't end up being the only ones that are converted.

   > *Re gender neutral bathrooms. I went to a non-AAS meeting recently that created gender neutral bathrooms by simply re-assigning women's rooms. I was very disappointed by this as it reduces the amount of space available to women even if they are fully utilizing the gender neutral space. This is particularly an issue since women are more likely to bring children to a meeting. Make sure either all bathrooms are converted to gender neutral or an equal number of men's and women's rooms.*

2. There was one comment in this pre-COVID survey that highlighted the potential of virtual conferences as a way to make them more equitable.

3. There were multiple comments highlighting the need to properly support the elevation of minoritized individuals into places of leadership. This support should come not only in the form of training and mentorship, but also in terms of compensation. At the same time, comments pointed out that it is not a valid assumption that minoritized individuals should be taking on the burden of fixing problems.

   > *DO NOT assume those who are most affected by a lack of equity and inclusion will be the ones to spend time and effort fixing the problems. We are already doing more than our fair share of service work and emotional labor every single day and we are tired and burned out.*

4. There were multiple comments pointing out that trainings were "preaching to the choir" and that they weren't reaching the people who most needed them, or who would find them most effective. As a counterpoint, however, one comment pointed out that trainings based on discussion were an effective way to improve the knowledge of those who were willing to learn, but not adequately informed.

5. There were multiple comments that the AAS should change the culture through clear actions ("walk the walk") as opposed to simply talking about change ("talk the talk").

   > *It is one thing to promote equity and inclusion, but entirely different to demonstrate it. If AAS is serious, they should pay close attention to their own actions, distribution of committees etc, and "walk the walk" from the top down.*

6. There were some reservations about forcing people to adhere to a set of norms.

   > *If this is forced onto people, there will be resentment and judgment, unfortunately. Again the problem is societal and trying to force people to adhere to some guideline will make honest people frustrated with the extra work and justifications for any hire and make prejudice people more prejudice.*

7. There were a few comments that stated a reluctance for the AAS to pursue Equity and Inclusion efforts. One of these statements felt that the leadership of the society was already diverse enough. Another claimed that the AAS should be a discussion venue for Equity and Inclusion work, but that it should not take any actions.

> *The AAS meetings have had some seminars and colloquia on harassment and inclusion in the past. The division of resources seems good now. To put more resources (or more scheduled time in) at the expense of science/results colloquia, however, would diminish my desire to attend AAS conferences. We need some, but probably not more than we have (unless they are parallel sessions, which are always a welcome option).*

## 4.3 Professional Development and Retention

### 4.3.1 Steps to support professional development

1. Award nominations and evaluations (and associated processes) should be transparent, but another comment indicated that revamping awards or developing awards for, e.g. mentoring, would only reward people who are accomplished and who don't actually need the reward.

   > *the AAS should follow best practices for evaluating awards, grants, etc as a model for the rest of the community*

   > *So much of the suggestions here - especially awards - are really going to play out in the rich-get-richer fashion. Someone who already has the network/stature to get a mentoring award is going to be well beyond needing it. Any actually worthwhile efforts will need to go after those who don't know how to help themselves.*

2. There was general consensus that the AAS should provide more training opportunities, at every step of the career path, from graduate school applications to negotiating salaries and start-up packages, in order for applicants to maximize their chances of being hired by the institution of their choice.
3. There is the need, however, to respond directly to the need of "minority scientist[s]" instead of assuming that those in the majority know best. For example, it is important to respect a diversity of career goals and support the many technical staff, support staff, and research astronomers who are not on the tenure track, but who still make fundamental contributions to the discipline. These individuals are often denied opportunities and resources for career development.

### 4.3.2 Steps to address hiring and the two-body problem

1. Given that so many astronomers are partnered with other astronomers (and/or in similar types of fields), the two-body problem is a pervasive and long-standing issue that needs to be addressed. Several of the solutions that were presented in the survey put the burden of solving the two-body problem on the people themselves

   > *limit opportunities to large cities where multiple opportunities exist*

   > *Accept work in separate locations permanently*

2. Others felt that the institutions should offer more opportunities for coupled astronomers.

   > *Departments should not be so narrowly focused on certain specialties. If researchers/faculty could have more freedom in where they could get a job then it would be easier for them and their families to find work in the same place.*

   > *Some institutions have adopted proactive policies; many have not. All too often it boils down to the people involved (e.g., is the hiring chair/dean a positive proactive person? or an old school sexist? or just a timid, lazy shite?)*

   > *Our department has lost out on 7 of the last 10 women we've tried to hire in the past decade because the University did not make a position available for a trailing academic partner.*

   > *I would venture to guess that a shared position is not desirable for most but I know of a couple that share a professorship and that has worked really well for them because they stuck to the half time aspect of it (the worry is that a shared position means the two individuals still end up working more than half the time but only get paid for half the time). To me it is appealing and I would like to see it as an option. Our culture is very career driven though so it might be dismissed easily from both sides.*

3. Clearly, the two-body problem affects people - and especially those who identify as women - profoundly, and the general feeling is that institutions and their departments should be encouraged to have policies in place to accommodate dual-career couples.
   - foregoing/delaying advertising a position if there is no possibility of a 2nd hire.
   - enabling joint hires, including cross-disciplinary ones.
   - creating second positions for the spouse/partners of the first hire, especially when the partner is a member of a historically marginalized group. Include assessments of such policies in the site visit reports.
     > *With the great number of purely-online programs, maybe it's reasonable for institutions to start reserving 'online-only' professorships for spousal hires. Of course, this is a problem if you have an application who is an online education expert that is not the spouse of a primary hire.*

4. Institutions should pay special attention to making accommodations for graduate students and postdocs, who are at critical stages in their careers.

   > *Remote postdoctoral positions are critical for those with a two body problem. Encourage all postdoctoral supervisors to allow it and encourage them to include extra money for travel so that postdoc and supervisors can have face to face interactions. For a permanent position it's easier to negotiate moving but it's very difficult to convince a partner to move every 2-3 years.*

5. Though at least one person commented that the AAS should not waste its resources *on developing a "recommendation",* multiple comments indicated that the AAS should take a more active role in supporting dual-career couples. The actions listed above work for industry and academia, and Astronomy in particular, would benefit from similar practices.

- include a mandatory/encouraged section of the AAS Job Register that describes more information about an institution's policy on dual hires
- recognize/reward institutions that negotiated a solution to the two-body problem
  > *If AAS doesn't provide tangible incentives, departments aren't going to do it.*

6. adopt a recommendation for best practices.
   > *Reasons to adopt positive two-body solution policies and what those policies are can be backed up by relevant literature.*

7. establish a postdoc-focused task force
8. *Be inclusive of the LGBTQ+ community in these discussions/remedies.*

### 4.3.3 Steps to support retention

1. Power issues need to be addressed
   > *There have to be practices that recognize bad behaviour which is often at the core of discouragement and includes all the obstacles that anyone (especially minorities) faces. What can be done to discourage those scientists, the ones that create the obstacles, from continuing on such an unproductive and offensive path?*

4. Funding and benefits offered by employment institutions need to be equitable, flexible, and long-term.
   > *Put pressure on funding agencies to work on developing funding initiatives and programs that emphasize flexible, long-term funding with employee benefits. The most pernicious and discriminatory problem in academia is the FUNDING STRUCTURE that rewards lots of cheap labor (multiple post-docs), and focuses on flashy new things (way more money to new observatories rather than researchers) to the detriment of long-term (stable) thought-out research programs. The main thing families can't handle and/or are less willing to tolerate is financial instability and uncertainty!*

5. There was a general consensus that employers need to develop policies that support people's caregiving responsibilities. There were multiple comments in support of extended paid parental leave and for institutional support for childcare. One idea to help caregivers, including those who care for sick and elderly family members, would be to offer them grants or fellowships to assist their return to the workforce.
   > *The time women have children is the unsafest time in a scientist's career. There should be a paid parental leave and policies in place to prevent their careers from being harmed. Support in finding affordable childcare is key.*

   > *There should be support given to scientists who want to get back into work after spending time as a care giver for a sick or elderly family member. This supports could be in the form of fellowships and*

> *grants similar to those who were taking care of new born born.*

## 4.4 Ethics

### 4.4.1 Comments on evaluating the effectiveness of different strategies that the AAS could pursue to improve professional ethics in the workplace

1. There was a suggestion that there be a cross-agency process to re-evaluate funding privileges when there is unethical behavior.
2. There were a number of recommendations for the AAS around potential policy actions. There was no consensus about what was appropriate for the AAS to do with regards to unethical behavior from favoring strong actions to saying that it was not in the AAS's purview to be investigating and instead have a series of reprimands dependent on the severity of the unethical behavior.

    > *If publication privileges are withdrawn, there should be clear steps and timelines for regaining those privileges. And we must ensure that we don't merely push papers that belong in AAS journals to other journals or other communication channels.*

    > *Perhaps consider the option of potentially penalizing entire departments or institutions that egregiously fail to protect marginalized members of its community from known harassers or other unsavory characters. This might make universities more willing to kick out, e.g., tenured professors with great publication records who have high-ranking advocates insisting allegations against them must be false.*

    > *Surveys don't mean anything if you aren't going to do anything with them. Also, having made ethics complaints to the AAS and been met with the email equivalent of a shrug and a "well what do you expect us to do?", I would say that the current structure and composition of the AAS has no interest in pursuing any violations of ethics. Of course....another element of this is also the journals: we should not for example have a single editor of any given science area, that just encourages a disguised system of patronage.*

3. Overall there was support for providing education on ethical issues and that the AAS should lead by example.

    > *Develop a curriculum of astronomy-specific ethical scenarios suitable for discussion.*

    > *Education is needed about what is sexism, racism etc. Different cultural circles understand it differently. For example, in the french culture a calendar on the office wall about naked women is not considered sexist. When we are all together on meetings, I want to be in such an environment, when sexist/racist/etc comments mean the same thing for all the colleagues and not based on their home culture.*

4. There was concern expressed around what actually is considered misconduct and whether or not that includes non-academic misconduct.

   > *AAS should focus on academic misconduct when discussing or promoting ethics. Going beyond that demonstrates internal biases.*

   > *Professional misconduct and scientific/academic dishonesty need to be kept separate. This science is not a social one, and shouldn't be made into one. We need responses to professional misconduct (racism, sexism, discrimination, etc.), but infractions of these rules need to be kept distinctly separate from infractions in research (stealing data, plagiarism, stealing credit, etc.). Astronomers are no more equipped to police human behavior than any other group, and shouldn't attempt to do so from behind a shield of "scientific professionalism." It is disingenuous.*

There was also some concern that the focus was on early-career researchers.

> *…there are far more older scientists who suffer from discrimination than there are transgendered scientists. And you just never mention ageism in this entire questionnaire. Shame on you!*

> *Remember that mid- and late-career people also had an early career and during that early career they may have had work stolen as well (or other impact). Why only consider interviewing early career people on this subject? Interviewing older people can show an even larger toll this took on their careers.*

5. There was overall consensus that rather than publicly calling out perpetrators, the emphasis should be on "calling in" to educate the community and to help people recognize and adapt their behaviors. There was also a sense that first-time offenders should be approached privately, while consistent and egregious behavior may require a more public solution.

   > *It's super-important to know how to respond to people engaging in *isms whether they realize it or not. The SBI-I Feedback Model (https://www.ccl.org/articles/leading-effectively-articles/closing-the-gap-between-intent-and-impact/) is an *AWESOME* way to do this. This has been really helpful in discussions in my department. I suggest that one person confront the perpetrator privately first. If they understand the problem, many times they'll respond and apologize right away. (Crisis averted.) If they don't, then have a second person talk to them. If they still don't listen, then address it publicly, with those affected by it.*

6. *Last two questions: develop a "calling in" culture, rather than a "calling out" culture. Goal should be to help people recognize and make amends for their own behavior, rather than public humiliation. (Unless it's repeated behavior, in which case, humiliate away!)*

However, there was some dissent.

> *Response to AAS membership engaging in racism, sexism, heterosexism, cissexism and/or ableism should be \*public\*, authoritative, and very clear.*

6. There was overall consensus about the importance of developing a culture of mutual respect that will enable open discussions of issues.

   > *I doubt dynamic situations are unambiguous in most cases. What is said (by one person) and heard (by another) often differ. I wonder if it might be better to cultivate a culture of mutual respect, rather than focusing on calling out wrongs and setting them right (although I am not saying we should not do that as well). Some of the harshest critics of AAS members' behavior are themselves often disrespectful of their colleagues' ideas and thoughts. This just breeds backlash, distrust and polarization. I believe you have to meet people where they are, not where you think they should be (even if you are right about where they should be). This takes work, not (just) Tweets.*

   > *Encourage departments to have open discussions about this. It seems like it happens everywhere and it is always kept hidden. A better way to fix the issues is to bring them out into the open if it is ok with the victims.*

7. There was some concern expressed that recognition of individuals can actually be counterproductive.
   > *Giving pats on the back out to people for doing good should only provide incentive to dishonest immoral people who lack any sort of integrity.*

   > *I do not like awards as they single out one good person over many equally good people. Responsible behavior should be acknowledged whenever it occurs.*

### 4.4.2 Other comments on professional ethics

1. There was support for ensuring that ethics violations in terms of publications should be addressed, and that there should be policies associated with that. There was associated concern that the AAS must carefully consider what falls within its purview to investigate and address and what does not.
   > *Allow victims to report retrospectively. Maybe an important paper was led by the supervisor instead of the student who did the work many years ago. There were no rules against unethical behavior at that times, but there is now, so the former student can report and reclaim their work which can still help on their career advancement. What I mean is try to prevent Jocelyn Bell cases, even retrospectively.*

   > *When there are issues that fall outside of the direct purview of the AAS (conduct at meetings, or journal issues), the issues should be handled by the home institution of the individuals involved. The AAS can, and should, provide support to the victims of alleged misconduct. But the AAS role in investigations in such cases should be limited. Different story for AAS Meetings and journal issues - there, the responsibility is squarely with the AAS. The AAS has a duty to be fair: always investigate, but*

> *do not assume automatically that every allegation is true, until reviewed. Possible misunderstandings, lack of information, or context, should be considered as factors in evaluating motive for an allegation.*

> *It would be helpful to end the practice of dishonest citing of previous work - but that seems a challenge. However, the AAS could compile the ratio of downloads to cites for papers in AAS journals and note whether some women/minorities have uncommonly large ratios of downloads to cites*

> *The AAS should be extremely careful that it doesn't try to set-up a private legal system within itself. We're not lawyers and judges and we shouldn't try to be. While I support strong penalties for misconduct, the AAS needs to be extremely careful in setting up policies and systems with as small a footprint as possible.*

2. Some thought that in addition to investigating potential misconduct, there should also be more proactive efforts to prevent it.

   > *Make the support systems for students more robust. There will always be a power structure and most people still do not officially report violations.*

3. A common thread was the need for consequences for inappropriate behavior.

   > *When people act unprofessionally, steal work, harass, bully, ect. Have real world consequences.*

   > *Sexual harassers should not be allowed to teach or attend AAS meetings until they have in some significant way acknowledged their error.*

   > *How to respond depends on what the unethical behavior is, how serious etc. There is a lot of perception here that can be highly variable.*

4. Some people pointed out that the impact on the victim of harassment needed to be considered.

   > *Responses to unethical behavior should not only be looked at in the context of the career of the individual being accused. They should also be looked at in the context of the individual that is the victim. How has their career been affected already by this unethical behavior?*

5. A number of people noted how important definitions are, but there was no consensus on whether harassment should be considered scientific misconduct.

   > *Definitions are important. Professional ethics and harassment have both similarities and differences.*

   > *Please do all of these things. The AGU has already defined harassment as scientific misconduct, and it's IMPERATIVE that we do the same! Science is more than the code on your screen, it's a human endeavor and we need to start recognizing that.*

   > *Harassment is not scientific misconduct. They are both serious issues, but should not be conflated.*

## 4.4.3 Additional feedback on ethics

1. There were comments on the need for resources.

   *I think it is important to develop an online guide and/or workshop to help people identify harassment, and give them guidelines as to what to do in a variety of situations — covering how to deal with a harasser on a day-to-day basis, and how and when it is appropriate to report them to your institution or law enforcement (or both).*

   *A single web page containing links to all the articles and papers written that explain clearly to perpetrators of unethical conduct, casual sexism etc etc why their behavior is unacceptable. This can also include articles on best practises e.g. how to write a reference letter for a woman etc. I want a website I can send these people to so they can do their homework rather than me having to explain it all to them. I don't have the time.*

2. There were a number of comments about the culture of the astronomical community, including the membership of the AAS.

   *I think abuse of power is a more generic issue than mistreatment of women and minorities, and it might be useful to widen the lens to include it.*

   *AAS has many internal political biases that should bring shame to a professional society.*

   *I think the AAS is doing a good job in identifying and combating harassment. I would like to make sure that transparency and process are promoted, so that men are not afraid to be in the same room with or work with women.*

   *Astronomers should be tolerant of the historical evolution of religious views.*

3. There were a few comments related to strategies to tackle issues related to ethics.

   *I think it is important for the AAS to coordinate with international bodies and societies in other countries on these issues. Our field is a global one, and many early career people work in multiple countries as students/postdocs. 'Fixing' these issues in the US will not solve the problem.*

   *I'm not clear on how this survey will inform a decadal white paper. This survey aises serious red flags about how the AAS is attempting to tackle large societal issues with only internal expertise (or lack thereof). Please consult external experts on issues the community prioritizes (you \*can\* survey community priorities).*

   *Fire who cause harassment, and hire the victim. Diversity is easy to check if it is applied in a group. Don't tolerate even a small fraction of gap in salary between men and women, etc... It seems to me that nothing has changed in astrophysics, something is even getting worse...*

4. A couple of people shared personal experiences.

   *I feel compelled to share that I, a senior woman, am being actively forced out this year by my institution and when I went to the AAS to find help, I found very little useful on the websites. I will likely be out of academia for good by this summer. This is something I have been fighting my whole career, and the worst offenders to me, personally, have always been the deans. Four terrible deans in three universities. I am done.*

   *There is no real mechanism in place to prevent harassment and discrimination from the AAS. This sort of behavior occurs in subtle ways. Even the AAS Homepage for the St. Louis meeting is discriminatory. If you are a minority, you can't win and have to be fighting all the time to get anywhere. Having surveys and "feel good" session does nothing. The AAS should hire a lawyer that its members can go to get advice and help with individual cases. That might actually help.*

5. One person brought up the issue of childcare specifically.

   *Childcare should not be emphasized as a women's issue. Caregiving is a societal issue for all genders and includes children, partners, and parents. Treat it as such and give no special treatment to mothers (this promotes sexism). We are all caregivers at some point in life and all need equal support. If childcare is offered at meetings, also include eldercare.*

6. Several actions the AAS and CSWA could take were suggested, while others were against taking action.

   *Transparency in how AAS committees (including CSWA) are populated would be nice-- can the AAS leadership issue an annual statement describing the pool of nominees from which people are selected? Many faculty searches are required to do this…*

   *I think the AAS CSWA should investigate the awarding of NASA grants, NSF grants , telescope time to women and minorities ... demonstrated biases exist in HST, NOAO, NRAO, "ESO, and Canada... this has a major effect on the professional careers of women. I think many/most of the 'suggestions' in this questionnaire are beyond the financial and personnel abilities of the AAS and the CSWA. And the ideas that the AAS CSWA is going to visit/inspect/confront all the astronomy-granting institutions to make them deal more 'inclusively' etc. is not a realistic plan.*

   *I find that many of the committees that work with these issues are echo chambers, and extremely reluctant to even listen to views of others. It is, I suppose a reflection of the times we are living in, but that is extremely offputting.*

   *Unfortunately this effort is dangerous and misguided. The AAS has limited resources to promote astronomy and this effort sounds like it wants to create a AAS policeforce that will be by its nature politcal and perhaps abusive. It may do some good, but it will surely do harm. Please think about what you're thinking about doing and reconsider.*

> *Get back to science and support scientists*

> *The society is currently falling apart, largely due to the efforts of committees such as this one. Society needs to return to focus on the scientific work of Astronomy.*

7. A few people brought up concerns about the survey itself.

> *Please talk to actual experts in these topics! I don't think my answers will be very useful because the only thing I can really judge is the merit of the goals of these policies, so I just end up saying most things are "very important."*

> *I am greatly disturbed by the biases inherent in this questionnaire. It is no surprise that it was written by a committee on the "status of women" in astronomy, since it gives minimal attention to categories of people more discriminated against in the AAS like racial minorities, the disabled, and older scientists. There is a long history of discrimination against women in society and in the sciences in particular. And in astronomy and planetary science. But astronomy has historically had a larger fraction of women in it than the other physical sciences beginning nearly a century ago. And, in planetary science in particular, substantial equity for women has been largely achieved in the last decade or two, although issues certainly remain. But there is blatant discrimination in astronomy against older scientists, in hiring decisions, in funding decisions, and even in minor issues like choosing chairs of sessions at AAS meetings. They are often under the guise of "encouraging young scientists." But it is blatant age-discrimination and it should be given at least equal attention. Where is the "Committee on the Status of Older Astronomers in Astronomy"? Where is the "Committee on the Status of Men and Women of Color in Astronomy"?*

> *Although I think the intentions of this survey are very positive, the execution seems to be a great example of how not to conduct an unbiased survey. Most of the questions beg the answers and try (however unintentionally) to guilt the respondent into how to answer. Wording is key here and there is a huge body of research behind how to write surveys of this nature to acquire data that is useful. Next time, the CSWA ought to get a bunch of experts (people in social anthropology) to review something of this nature before sending it out, or better to write it based on the research on how to limit bias in the data that comes out of a survey. The results of this, however interesting, will certainly be questionable, I imagine.*

## 5. Summary

The CSWA conducted a large community survey to inform a set of recommendations to the AAS. This paper - Paper I of a two-paper series - outlines the background of the status of women in astronomy (§2), the methods used to collect the data using the survey (§3), and the data collected by this survey (§4). The data include both quantitative information from Likert-scale questions and qualitative responses from free-response questions. In §4 we summarize the free-response questions related to four general focus areas: harassment & bullying (§4.1), creating inclusive environments (§4.2), professional development and retention (§4.3), and ethics (§4.4). The quantitative results are given in the Appendix, and the full set of qualitative responses as well as a PDF of the survey can be found in an on-line repository(Zellner, Knezek, Wexler, & Committee on the Status of Women in Astronomy, 2022). There were over 340 responses to the survey and the recommendations informed by these data are presented in Paper II.

There was a broad spectrum of input from the community, and in summarizing the themes touched on by the qualitative responses we present synopses of opinions with accompanying example comments that attempt to illustrate the range of responses in any given area. In some areas there was no clear consensus, but in others the opinions of the community were made very clear. The input of the community was invaluable in crafting our recommendations and the large number of responses ensures that the recommendations in Paper II represent a cross-sectional view of the CSWA constituency.

The purpose of this series of papers is specifically to present recommendations to the AAS. However, the free-responses in the survey sometimes touched upon issues beyond the purview of the AAS and called out actions that can be taken by institutions or funding agencies. While we present all the data here, our recommendations in Paper II focus on those items within the sphere of the AAS. Data from this survey were also used to inform a set of white papers presented to the Astro2020 Decadal Survey, whose target audiences are the federal funding agencies (Zellner et al. 2019a, 2019b).

# Acknowledgements

Wexler thanks NASA for its support through the NASA internship program. The authors thank the astronomical community for their responses to the survey, which directly enabled this work. Rudnick thanks the NSF for their support through AAG grants AST-1716690, 2206473, and 2308126.

# Appendix - Quantitative survey results

Here we include the quantitative survey results for the Likert-scale opinion questions broken down into the four focus areas of the survey [3]. For each survey item the percentages reflect the percentage of respondents for that question who indicated each of the choices. We only report integer percentages and because of rounding not all rows will add up to 100%. Each table is preceded by the prompt given to the survey respondents.

## A.1 Harassment and Bullying

Please evaluate the likely effectiveness of the following strategies that the AAS could pursue in order to prevent harassment.

| Table 1 |
|---------|

|  | Not at all effective | A little effective | Somewhat effective | Very effective | No opinion |
| --- | --- | --- | --- | --- | --- |
| Enforce Anti-Harassment Policies | 4% | 8% | 40% | 43% | 5% |
| Create an Information Escrow | 8% | 10% | 36% | 33% | 12% |
| Encourage departments and grad schools to include anti-harassment training both in their orientation activities and as part of their university policies | 7% | 17% | 36% | 38% | 2% |
| Encourage bystander intervention training | 6% | 11% | 31% | 47% | 4% |
| Encourage institutions to develop anti-harassment training that is customized to specific departments or locations | 8% | 22% | 36% | 28% | 7% |
| Hold perpetrators accountable and provide public, anonymized information about completed investigations | 6% | 5% | 21% | 61% | 8% |

How should the AAS support those who may be or have been harmed by harassment and bullying?

| Table 2 | | | | | |
|---|---|---|---|---|---|
| | Not at all important | A little important | Important | Very Important | No opinion |
| Provide mentoring and/or counseling for those who have been adversely affected by harassment | 5% | 17% | 30% | 43% | 5% |
| Teach people how to speak up and advocate for themselves in the case that they are harassed or bullied | 6% | 17% | 30% | 45% | 2% |
| Support people in deciding whether to file a complaint with the AAS Percentage | 4% | 9% | 28% | 54% | 4% |
| Support people in deciding whether to file a complaint with their home institution or other organizations. | 5% | 9% | 24% | 58% | 4% |

## A.2 Creating Inclusive Environments

Please evaluate the likely effectiveness of the following strategies that the AAS could pursue in fostering equity and inclusion across the board, especially for community members with intersectional marginalized identities

| Table 3 | | | | | |
|---|---|---|---|---|---|
| | Not at all effective | A little effective | Somewhat effective | Very Effective | No opinion |
| Educate members about ways to make diversity and inclusion a priority, not an afterthought | 5% | 14% | 39% | 40% | 2% |
| Educate members on preventing reverse discrimination that may occur inadvertently in the pursuit of inclusivity | 20% | 21% | 26% | 26% | 8% |
| Work with SGMA, WGAD, CSMA, CSWA, and others, to promote gender-neutral bathrooms, lactation rooms, and other provisions for marginalized groups | 9% | 10% | 28% | 45% | 7% |
| Encourage universities to designate faculty members to support equity and inclusion issues not applicable to administrative offices such as Title VII / IX. | 13% | 16% | 31% | 29% | 11% |
| Encourage institutions to provide universal access to their facilities | 7% | 16% | 21% | 42% | 14% |
| Continue to provide support for regional meetings that focus on diversity, equity, and inclusion issues | 9% | 18% | 26% | 41% | 6% |
| Continue to provide support for sessions that focus on diversity, equity, and inclusion issues at AAS meetings | 7% | 13% | 27% | 49% | 4% |
| Provide funding that increases the accessibility of networking and meeting spaces | 4% | 13% | 31% | 46% | 6% |
| Schedule conferences, seminars, and meetings at family-friendly times, be flexible when scheduling events, and provide video conferencing capabilities | 3% | 6% | 26% | 61% | 4% |
| Increase equitable access to policy making and leadership roles within the AAS and its divisions; deliberately reach out to and involve individuals from across the entire astronomical community, especially underrepresented and under-resourced researchers and institutions, in policy and leadership roles. | 4% | 9% | 24% | 58% | 5% |

## A.3 Professional Development, Hiring, and Retention

Please evaluate the likely effectiveness of the following strategies that the AAS could pursue to improve professional development, hiring, and workplace environments

| Table 4 | | | | | |
|---|---|---|---|---|---|
| | Not at all effective | A little effective | Somewhat effective | Very effective | No opinion |
| Offer a mentoring program for astronomers in all career stages | 5% | 15% | 40% | 37% | 4% |
| Promote mentoring best practices | 2% | 13% | 36% | 46% | 3% |
| Develop a salary database to support efforts toward equity in pay | 6% | 14% | 23% | 53% | 4% |
| Support astronomers at all career stages from small institutions or non-academic organizations who may not have access to the same support network as those at larger institutions | 3% | 9% | 32% | 52% | 4% |
| Encourage institutions to implement ways to mitigate implicit bias in the workplace (including student settings) | 7% | 16% | 34% | 41% | 1% |
| Enable greater diversity (gender, ethnic, racial, geographical, institutional, etc.) in the pool of nominees for AAS and Division prizes and committees, and capture the data on the pools as a function of time | 8% | 9% | 27% | 51% | 5% |
| Create a mentoring award | 13% | 22% | 29% | 29% | 6% |
| Facilitate leadership training (e.g., financial skills, high-level program management, personnel skills) for AAS members | 6% | 14% | 37% | 38% | 5% |
| Adopt best practices to encourage women and other members of marginalized groups to pursue leadership positions | 4% | 8% | 33% | 51% | 4% |
| Offer more resources such as workshops to combat imposter syndrome | 12% | 16% | 34% | 33% | 5% |
| For all career levels, provide funding and access to resources to mitigate the extra impact of caregiving on women | 6% | 10% | 26% | 55% | 4% |
| Encourage institutions to work to mitigate the two-body problem | 10% | 15% | 25% | 45% | 6% |
| Continue to encourage and provide opportunities for instructors, potential instructors, and teaching assistants to learn new pedagogical and assessment techniques (i.e. workshops, mentoring for teaching) | 5% | 19% | 34% | 36% | 6% |
| Provide incentives and opportunities (such as awards, grants, and workshops) for instructors to develop and/or implement research-based inclusive teaching practices | 7% | 17% | 32% | 38% | 6% |
| Make dual-anonymous refereeing of papers mandatory | 18% | 13% | 20% | 39% | 9% |

If you have experience with the two-body problem, please evaluate the effectiveness of the following approaches where applicable.

| Table 5 | | | | | |
|---|---|---|---|---|---|
| | Not at all effective | A little effective | Somewhat effective | Very effective | Not applicable |
| Be flexible - work from home | 7% | 13% | 35% | 45% | 0% |
| Be flexible - accept work in separate locations temporarily | 17% | 22% | 32% | 29% | 0% |
| Ask for compressed teaching schedules | 21% | 19% | 35% | 25% | 0% |
| Negotiate a shared position | 29% | 26% | 21% | 24% | 0% |
| Negotiate a second position | 9% | 12% | 16% | 62% | 0% |

Rate the importance of the following strategies that the AAS can adopt to help alleviate the two-body problem.

| Table 6 | | | | | |
|---|---|---|---|---|---|
| | Not at all Important | Somewhat Important | Important | Very Important | No Opinion |
| Encourage institutions to enable remote work | 7% | 21% | 29% | 33% | 10% |
| Solicit personal accounts from members on their experiences with the two-body problem | 10% | 28% | 27% | 22% | 13% |
| Find ways to encourage/mentor those in that situation | 9% | 23% | 29% | 26% | 12% |

## A.4 Ethics

Please evaluate the likely effectiveness of the following strategies that the AAS could pursue to improve professional ethics in the workplace.

| Table 7 | | | | | |
|---|---|---|---|---|---|
| | Not at all effective | A little effective | Somewhat effective | Very Effective | No opinion |
| Survey early-career scientists about whether they have experienced academic dishonesty/having their work stolen by more senior, scientists and, if so, how the experience has affected their careers | 6% | 17% | 39% | 37% | 1% |
| Define harassment and/or other unethical behavior as a form of scientific misconduct and invoke a similar enforcement process as for scientific misconduct | 7% | 9% | 25% | 57% | 3% |
| Withdraw AAS privileges, such as publishing in its journals, in response to unethical behavior | 8% | 9% | 16% | 62% | 4% |
| Remove AAS honors, such as prizes and awards, in response to unethical behavior | 9% | 8% | 16% | 64% | 3% |
| Enforce the AAS Code of Ethics according to the provisions in the code | 3% | 7% | 23% | 60% | 7% |
| Support astronomers in learning to recognize and acknowledge their responsibility to be 'good citizens' in areas where their research interacts with societal concerns | 5% | 14% | 37% | 40% | 4% |
| Provide an award for good service to society and/or good citizenship in the above sense | 18% | 25% | 29% | 23% | 5% |
| Support the AAS leadership in learning to be models for good citizenship in the above sense | 10% | 20% | 38% | 27% | 5% |
| Respond promptly when astronomers publicly engage in racism, sexism, heterosexism, cissexism, and/or ableism during meetings sponsored by the AAS and its divisions | 5% | 7% | 20% | 67% | 1% |
| Develop a process for how/if to respond when astronomers publicly engage in racism, sexism, heterosexism, cissexism, and/or ableism during activities not specifically sponsored by the AAS and its divisions. | 6% | 9% | 19% | 63% | 2% |

# Footnotes

1. During the development of the original survey the term "members of underrepresented groups (URM)" was commonly accepted as an appropriate term. A more recent consensus is to use "members of historically marginalized groups". We refer to the latter throughout this paper, though original survey questions and comments quoted verbatim still adopt the original term. ↩

2. We note that sometimes in the surveys referenced here that women sometimes is defined to solely encompass women and sometimes it encompasses women and gender minorities. ↩

# References

- Aloisi, A., & Reid, N. (2021). (Un)conscious Bias in the Astronomical Profession: Universal Recommendations to improve Fairness, Inclusiveness, and Representation. *Bulletin of the AAS*, *53*(4). https://doi.org/10.3847/25c2cfeb.299343eb